\documentstyle[prl,eqsecnum,aps]{revtex}

\begin{document}
\author{Jian-Qi Shen $^{1,}$$^{2}$ \footnote{E-mail address: jqshen@coer.zju.edu.cn}}
\address{$^{1}$  Centre for Optical
and Electromagnetic Research, State Key Laboratory of Modern
Optical Instrumentation, Zhejiang University,
Hangzhou Yuquan 310027, P. R. China\\
$^{2}$Zhejiang Institute of Modern Physics and Department of
Physics, Zhejiang University, Hangzhou 310027, P. R. China}
\date{\today }
\title{Geometric phase due to helicity inversion of photons inside an optical fiber composed periodically of left- and right- handed media}
\maketitle

\begin{abstract}
In this Letter, it is claimed that, in addition to the Chiao-Wu
geometric phase and optical Aharonov-Carmi geometric phase, there
exists a new interesting geometric phase caused by the helicity
inversion of photons in the optical fiber composed periodically of
left- and right- handed (LRH) media. By making use of the
Lewis-Riesenfeld invariant theory and the invariant-related
unitary transformation formulation, we calculate this geometric
phase. It is emphasized that this geometric phase should not been
neglected when considering the anomalous refraction on the
interfaces between left- and right- handed media.
\\ \\
\end{abstract}

More recently, a kind of composite media ( the so-called {\it
left-handed media} ) having a frequency band where the effective
permittivity ($\varepsilon$) and the effective permeability
($\mu$) are {\it simultaneously negative} attracts attention of
many researchers in various fields such as materials science,
condensed matter physics, optics and
electromagnetism\cite{Smith,Klimov,Pendry3,Shelby,Ziolkowski2}. In
1964\footnote{Note that, in the literature, many authors mention
the year when Veselago suggested the {\it left-handed media} by
mistake. They claim that Veselago proposed the concept of {\it
left-handed media} in 1968. On the contrary, the true fact is as
follows: Veselago's excellent paper was first published in July,
1964 [Usp. Fiz. Nauk {\bf 92}, 517-526 (1964)]. In 1968, this
original paper was translated into English by W. H. Furry and
published again in the journal of Sov. Phys.
Usp.\cite{Veselago}.}, Veselago considered this peculiar medium
and showed that it possesses a {\it negative} index of
refraction\cite{Veselago}. It follows from the Maxwell's equations
that in this medium the Poynting vector and wave vector of
electromagnetic wave would be antiparallel, {\it i. e.}, the wave
vector {\bf {k}}, the electric field {\bf {E}} and the magnetic
field {\bf {H}} form a {\it left-handed} system; thus Veselago
referred to such materials as ``{\it left-handed} (LH) '', and
correspondingly, the ordinary medium in which {\bf {k}}, {\bf {E}}
and {\bf {H}} form a right-handed system may be termed the  ``{\it
right-handed} (RH) '' medium. Other authors call this class of
materials ``negative-index media (NIM)'', ``double negative media
(DNM) \cite{Ziolkowski2}'' and Veselago's media. It is readily
verified that in such media having both $\varepsilon$ and $\mu$
{\it negative}, there exist a number of peculiar electromagnetic
properties, for instance, many dramatically different propagation
characteristics stem from the sign change of the group velocity,
including reversal of both the Doppler shift and Cherenkov
radiation, anomalous refraction, modified spontaneous emission
rates and even reversals of radiation pressure to radiation
tension\cite{Klimov,Veselago}. In experiments, this artificial
negative electric permittivity media may be obtained by using the
{\it array of long metallic wires} (ALMWs), which simulates the
plasma behavior at microwave frequencies, and the artificial
negative magnetic permeability media may be built up by using
small resonant metallic particles, {\it e. g.}, the {\it split
ring resonators} (SRRs) with very high magnetic
polarizability\cite{Pendry3,Pendry1,Pendry2,Maslovski}.

All the above optical and electromagnetic properties and effects
of left-handed media were predicted and studied based on the
classical Maxwell's theory. Since left-handed media possess so
many peculiar properties and effects, we think it is also
physically interesting to consider some certain quantum effects in
the left-handed optical fiber. So, in this paper, we deal with a
problem of geometric quantum phase of photons inside the left- and
right- handed (LRH)- periodical fiber.

Historically, the first physical realization of the Berry's
quantum geometric (topological) phase\cite {Berry} was proposed by
Chiao and Wu\cite {Chiao,Tomita,Kwiat,Robinson} after Berry
suggested the adiabatic quantum theory. By using Berry's adiabatic
phase formula\cite {Berry}, Chiao and Wu investigated the
geometric phase of the polarized photon propagating inside a
noncoplanarly curved optical fiber\cite {Chiao}. Based on the
Lewis-Riesenfeld invariant theory\cite{Lewis,Gao1,Gao2}, we have
calculated the non-adiabatic non-cyclic Chiao-Wu geometric phase
of photons\cite{Shen1}. Since it is readily verified that the
photon propagating inside a curved fiber may be regarded as acted
upon by an effective Coriolis force ( Lorentz-type force
)\cite{Shen1,Zhu}, one can predict that there exists an optical
Aharonov-Carmi effect\cite{Anandan,Dresden} for moving photons in
the curved optical fiber, where the geometric phase is similar to
the Aharonov-Bohm phase.

In the present Letter, we argue that, if, for example, the photons
propagating inside a fiber that is composed periodically of left-
and right- handed (LRH) media, then a new geometric phase of
photons due to helicity inversion will arise. In this
LRH-periodical optical fiber, helicity inversion ( or the
transitions between helicity states ) of photons may be easily
caused by the interaction of light field with interfaces of LRH
materials. So, this geometric phase arises on the interfaces
between left- and right- handed media ( LRH interfaces ), where
the anomalous refraction occurs when the incident lightwave
travels to the LRH interfaces. we think, in the literature, it
gets less attention than it deserves. So, in what follows, we
calculate the photon wavefunction and geometric phase in this
physical process, and emphasize that we should attach importance
to this geometric phase when considering the wave propagation on
the LRH interfaces.

Let us consider the applied optical fiber that is fabricated
periodically from both left- and right- handed (LRH) media with
the optical refractive index being $-n$ and $n$, respectively.
Thus the wave vector of photon moving along the fiber is
respectively $-n\frac{\omega }{c}$ in LH section and
$n\frac{\omega }{c}$ in RH section, where $\omega$ and $c$
respectively denote the frequency and speed in a vacuum of light.
For convenience, we assume that the periodical length, $b$, of LH
is equal to that of RH in the fiber. If the eigenvalue of photon
helicity is $\sigma $ in right-handed sections, then, according to
the definition of helicity, $h=\frac{\bf k}{\left| {\bf k}\right|
}\cdot{\bf J}$ with ${\bf J}$ denoting the total angular momentum
of the photon, the eigenvalue of helicity acquires a minus sign in
left-handed sections. We assume that at $t=0$ the light propagates
in the right-handed section and the initial eigenvalue of photon
helicity is $\sigma $. So, in the wave propagation inside the
LRH-periodical optical fiber, the helicity eigenvalue of $h$ is
then $\left( -\right) ^{m}\sigma $ with $m=\left[
\frac{ct}{nb}\right] $, where $\left[ {\quad }\right] $ represents
the integer part of $\frac{ct}{nb}$. This, therefore, means that
when $2k\left( \frac{nb}{c}\right) <t\leq \left( 2k+1\right)
\left( \frac{nb}{c}\right) $, $\left( -\right) ^{m} =+1$ and when
$\left( 2k+1\right) \left( \frac{nb}{c}\right) <t\leq (2k+2)\left(
\frac{nb}{c}\right) $, $\left( -\right) ^{m} =-1$, where $k$ is an
integer ($k\geq 0$). It follows that the incidence of lightwave in
the fiber on the LRH interfaces gives rise to transitions between
the helicity states ( $\left| +\right\rangle $ and $\left|
-\right\rangle $ ) of photons. This enables us to construct a
time-dependent effective Hamiltonian
\begin{equation}
H\left( t\right) =\frac{1}{2}\omega \left( t\right) \left(
S_{+}+S_{-}\right)    \eqnum{1}                  \label{eq1}
\end{equation}
in terms of $\left| +\right\rangle $ and $\left| -\right\rangle $
to describes this transition process of helicity states on the LRH
interfaces, where $S_{+}=\left| +\right\rangle \left\langle
-\right| ,S_{-}=\left| -\right\rangle \left\langle +\right|
,S_{3}=\frac{1}{2}\left( \left| +\right\rangle \left\langle
+\right| -\left| -\right\rangle \left\langle -\right| \right) $
satisfying the following SU(2) Lie algebraic commuting relations
$\left[ S_{+},S_{-}\right] =2S_{3}$ and $\left[ S_{3},S_{\pm
}\right] =\pm S_{\pm }$. The time-dependent frequency parameter
$\omega \left( t\right) $ may be taken to be $\omega \left(
t\right) =\zeta \frac{\rm d}{{\rm d}t}p\left( t\right) $, where
$p\left( t\right) =\left( -\right) ^{m}$ with $m=\left[
\frac{ct}{nb}\right] $, and $\zeta$ is the coupling coefficient,
which is determined by the physical mechanism of interaction
between light field and media. Since $p\left( t\right) $ is a
periodical function, by using the analytical continuation
procedure, it can be rewritten as the following combinations of
analytical functions
\begin{equation}
p\left( t\right) =\sum_{k=1}^{\infty }\frac{2}{k\pi }\left[
1-\left( -\right) ^{k}\right] \sin \left( \frac{k\pi
c}{nb}t\right).    \eqnum{2}                  \label{eq2}
\end{equation}
In what follows, we solve the time-dependent Schr\"{o}dinger
equation ( in the unit $\hbar =1$ )
\begin{equation}
H\left( t\right) \left| \Psi _{\sigma }\left( t\right)
\right\rangle =i\frac{\partial }{\partial t}\left| \Psi _{\sigma
}\left( t\right) \right\rangle     \eqnum{3} \label{eq3}
\end{equation}
governing the propagation of light in the LRH- periodical fiber.
According to the Lewis-Riesenfeld invariant theory, the exact
particular solution $\left| \Psi _{\sigma }\left( t\right)
\right\rangle$ of the time-dependent Schr\"{o}dinger equation
(\ref{eq3}) is different from the eigenstate of the invariant
$I(t)$ only by a time-dependent $c$- number factor $\exp \left[
\frac{1}{i}\varphi _{\sigma }\left( t\right) \right]$, where
\begin{equation}
\varphi _{\sigma }\left( t\right)=\int_{0}^{t}\left\langle \Phi
_{\sigma }\left( t^{\prime }\right) \right|[H(t^{\prime
})-i\frac{\partial }{\partial t^{\prime }}]\left| \Phi _{\sigma
}\left( t^{\prime }\right) \right\rangle {\rm d}t^{\prime }
\eqnum{4} \label{eq4}
\end{equation}
with $\left| \Phi _{\sigma }\left( t\right) \right\rangle $ being
the eigenstate of the invariant $I(t)$ ( corresponding to the
particular eigenvalue $\sigma$ ) and satisfying the eigenvalue
equation $I\left( t\right) \left| \Phi _{\sigma }\left( t\right)
\right\rangle =\sigma \left| \Phi _{\sigma }\left( t\right)
\right\rangle$, where the eigenvalue $\sigma$ of the invariant
$I(t)$ is {\it time-independent}. Thus we have
\begin{equation}
\left| \Psi _{\sigma }\left( t\right) \right\rangle =\exp \left[
\frac{1}{i}\varphi _{\sigma }\left( t\right) \right] \left| \Phi
_{\sigma }\left( t\right) \right\rangle.   \eqnum{5} \label{eq51}
\end{equation}
So, in order to obtain $\left| \Psi _{\sigma }\left( t\right)
\right\rangle$, we should first obtain the eigenstate $\left| \Phi
_{\sigma }\left( t\right) \right\rangle $ of the invariant $I(t)$.
In accordance with the Lewis-Riesenfeld theory, the invariant
$I(t)$ is a conserved operator ( {\it i. e.}, it possesses {\it
time-independent} eigenvalues ) and agrees with the following
Liouville-Von Neumann equation

\begin{equation}
\frac{\partial I(t)}{\partial t}+\frac{1}{i}[I(t),H(t)]=0.
\eqnum{6}                  \label{eq5}
\end{equation}
It follows from Eq. (\ref{eq5}) that the invariant $I(t)$ may also
be constructed in terms of $S_{\pm }$ and $S_{3}$, {\it i. e.},

\begin{equation}
I\left( t\right) =2\left\{\frac{1}{2}\sin \theta \left( t\right)
\exp \left[ -i\phi  \left( t\right)\right] S_{+}+\frac{1}{2}\sin
\theta \left( t\right) \exp \left[ i\phi  \left( t\right)\right]
S_{-}+\cos \theta \left( t\right) S_{3}\right\}. \eqnum{7}
\label{eq6}
\end{equation}
Inserting Eq. (\ref{eq1}) and (\ref{eq6}) into Eq. (\ref{eq5}),
one can arrive at a set of auxiliary equations

\begin{eqnarray}
\exp \left[ -i\phi \right] \left( \dot{\theta}\cos \theta
-i\dot{\phi}\sin \theta \right) -i\omega \cos \theta =0,  \nonumber \\
\quad \dot{\theta}+\omega \sin \phi =0, \eqnum{8} \label{eq7}
\end{eqnarray}
which are used to determine the time-dependent parameters,
$\theta\left( t\right)$ and $\phi\left( t\right)$, of the
invariant $I(t)$.

It should be noted that we cannot easily solve the eigenvalue
equation $I\left( t\right) \left| \Phi _{\sigma }\left( t\right)
\right\rangle =\sigma \left| \Phi _{\sigma }\left( t\right)
\right\rangle$, for the time-dependent parameters $\theta\left(
t\right)$ and $\phi\left( t\right)$ are involved in the invariant
(\ref{eq6}). If, however, we could find ( or construct ) a unitary
transformation operator $V(t)$ to make $V^{\dagger }(t)I(t)V(t)$
be {\it time-independent}, then the eigenvalue equation problem of
$I(t)$ is therefore easily resolved. According to our experience
for utilizing the invariant-related unitary transformation
formulation\cite{Shen2}, we suggest a following unitary
transformation operator

\begin{equation}
V\left( t\right) =\exp \left[ \beta \left( t\right) S_{+}-\beta
^{\ast }\left( t\right) S_{-}\right] , \eqnum{9} \label{eq8}
\end{equation}
where $\beta (t)$ and $\beta ^{\ast }(t)$ will be determined by
calculating $I_{\rm V}=V^{\dagger }(t)I(t)V(t)$ in what follows.

Calculation of $I_{\rm V}=V^{\dagger }(t)I(t)V(t)$ yields

\begin{equation}
I_{\rm V}=V^{\dagger }\left( t\right) I\left( t\right) V\left(
t\right) =2S_{3},  \eqnum{10} \label{eq9}
\end{equation}
if $\beta $ and $\beta ^{\ast }$ are chosen to be $\beta \left(
t\right) =-\frac{\theta \left( t\right) }{2}\exp \left[ -i\phi
\left( t\right) \right] ,\quad \beta ^{\ast }\left( t\right)
=-\frac{\theta \left( t\right) }{2}\exp \left[ i\phi \left(
t\right) \right] $. This, therefore, means that we can change the
time-dependent $I(t)$ into a time-independent $I_{\rm V}$, and the
result is $I_{\rm V}=2S_{3}$. Thus, the eigenvalue equation of
$I_{\rm V}$ is $I_{\rm V}\left| \sigma \right\rangle =\sigma
\left| \sigma \right\rangle $ with $\sigma=\pm 1$, and
consequently the eigenvalue equation of $I(t)$ is $I\left(
t\right) V\left( t\right) \left| \sigma \right\rangle =\sigma
V\left( t\right) \left| \sigma \right\rangle $. So, we obtain the
eigenstate $\left| \Phi _{\sigma }\left( t\right) \right\rangle$
of $I(t)$, {\it i. e.}, $\left| \Phi _{\sigma }\left( t\right)
\right\rangle=V\left( t\right)\left| \sigma \right\rangle$.

Correspondingly, $H(t)$ is transformed into

\begin{eqnarray}
H_{\rm V}\left( t\right) =V^{\dagger }\left( t\right) \left[
H\left( t\right) -i\frac{\partial }{\partial t}\right] V\left(
t\right) \eqnum{11} \label{eq11}
\end{eqnarray}
and the time-dependent Schr\"{o}dinger equation (\ref{eq3}) is
changed into
\begin{equation}
H_{\rm V}\left( t\right) \left| \Psi _{\sigma }\left( t\right)
\right\rangle_{\rm V} =i\frac{\partial }{\partial t}\left| \Psi
_{\sigma }\left( t\right) \right\rangle_{\rm V}  \eqnum{12}
\label{eq12}
\end{equation}
under the unitary transformation $V(t)$, where $ \left| \Psi
_{\sigma }\left( t\right) \right\rangle_{\rm V}=V^{\dagger }\left(
t\right) \left| \Psi _{\sigma }\left( t\right) \right\rangle$.

Further analysis shows that the exact particular solution $ \left|
\Psi _{\sigma }\left( t\right) \right\rangle_{\rm V}$ of the
time-dependent Schr\"{o}dinger equation (\ref{eq12}) is different
from the eigenstate $\left| \sigma \right\rangle$ of the {\it
time-independent} invariant $I_{\rm V}$ only by a time-dependent
$c$- number factor $\exp \left[ \frac{1}{i}\varphi _{\sigma
}\left( t\right) \right]$\cite {Lewis}, which is now rewritten as
$\exp \left \{ \int_{0}^{t}\left\langle \sigma\right| H_{\rm
V}\left( t^{\prime }\right)\left|\sigma \right\rangle {\rm
d}t^{\prime }\right\}$.

By using the auxiliary equations (\ref{eq7}), it is verified that
$H_{\rm V}(t)$ depends only on the operator $S_{3}$, {\it i. e.},

\begin{equation}
H_{\rm V}\left( t\right) =\left\{ \omega \left( t\right)\sin
\theta \left( t\right) \cos \phi \left( t\right)+\dot{\phi}\left(
t\right) \left[ 1-\cos \theta \left( t\right) \right] \right\}
S_{3}   \eqnum{13} \label{eq13}
\end{equation}
and the time-dependent $c$- number factor $\exp \left[
\frac{1}{i}\varphi _{\sigma }\left( t\right) \right] $ is
therefore $\exp \left\{ \frac{1}{i}\left[ \varphi _{\sigma
}^{\left( {\rm d}\right) }\left( t\right) +\varphi _{\sigma
}^{\left( {\rm g}\right) }\left( t\right) \right] \right\} $,
where the dynamical phase is

\begin{equation}
\varphi _{\sigma }^{\left( {\rm d}\right) }\left( t\right) =\sigma
\int_{0}^{t}\omega \left( t^{\prime }\right)\sin \theta \left(
t^{\prime }\right) \cos \phi \left( t^{\prime }\right){\rm
d}t^{\prime } \eqnum{14} \label{eq14}
\end{equation}
and the geometric phase is
\begin{equation}
\varphi _{\sigma }^{\left( {\rm g}\right) }\left( t\right) =\sigma
\int_{0}^{t}\dot{\phi}\left( t^{\prime }\right) \left[ 1-\cos
\theta \left( t^{\prime }\right) \right] {\rm d}t^{\prime }.
\eqnum{15} \label{eq15}
\end{equation}

Hence the particular exact solution of the time-dependent
Schr\"{o}dinger equation (\ref{eq3}) corresponding to the
particular eigenvalue, $\sigma$, of the invariant $I(t)$ is of the
form

\begin{equation}
\left| \Psi _{\sigma }\left( t\right) \right\rangle =\exp \left\{
\frac{1}{i}\left[ \varphi _{\sigma }^{\left( {\rm d}\right)
}\left( t\right) +\varphi _{\sigma }^{\left( {\rm g}\right)
}\left( t\right) \right] \right\} V\left( t\right) \left| \sigma
\right\rangle. \eqnum{16} \label{eq16}
\end{equation}

It follows from the obtained expression (\ref{eq15}) for geometric
phase of photons that, if the frequency parameter $\omega$ is
small ({\it i. e.}, the adiabatic quantum process ) and then
according to the auxiliary equations (\ref{eq7}),
$\dot{\theta}\simeq 0$, the Berry phase ( adiabatic geometric
phase ) in a cycle ( {\it i. e.}, one round trip, $T\simeq
\frac{2\pi}{\omega}$ ) of parameter space of invariant $I(t)$ is

\begin{equation}
\varphi _{\sigma }^{\left( {\rm g}\right) }\left( T\right)= 2\pi
\sigma (1-\cos \theta ), \eqnum{17} \label{eq17}
\end{equation}
where $2\pi (1-\cos \theta )$ is a solid angle over the parameter
space of the invariant $I(t)$, which means the geometric phase is
related only to the geometric nature of the pathway along which
quantum systems evolve. Expression (\ref{eq17}) is analogous to
the magnetic flux produced by a monopole of strength $\sigma$
existing at the origin of the parameter space. This, therefore,
implies that geometric phases differ from dynamical phases and
involve the global and topological properties of the time
evolution of quantum systems.

It is worthwhile to point out that the geometric phase of photons
due to helicity inversion presented here is of quantum level.
However, whether the Chiao-Wu geometric phase due to the spatial
geometric shape of fiber is of quantum level or not is not
apparent ( see, for example, the arguments between Haldane and
Chiao, Tomita about this problem \cite{Haldane2,Chiao2}), since
the expression for the Chiao-Wu geometric phase can be derived by
using both the classical Maxwell's electromagnetic theory,
differential geometry and quantum
mechanics\cite{Kwiat,Robinson,Haldane1,Haldane2,Chiao2}. However,
the geometric phase in this Letter can be considered only by
Berry's adiabatic quantum theory and Lewis-Riesenfeld invariant
theory, namely, the classical electrodynamics cannot predict this
geometric phase. Although many investigators have taken into
account the boundary condition problem and anomalous refraction in
left-handed media by using the classical Maxwell's
theory\cite{Shelby,Veselago}, less attention is paid to this
geometric phase due to helicity inversion. It is believed that
this geometric phase originates at the quantum level, but survives
the correspondence-principle limit into the classical level. So,
We emphasize that it is essential to take into consideration this
geometric phase in investigating the anomalous refraction on the
LRH interfaces. In addition, in the Chiao-Wu case, there exists
the vacuum geometric phase at pure quantum level, which arises
from the vacuum quantum fluctuations of photons field\cite{Shen1}.
We think such quantum vacuum geometric phase deserves
consideration also in the present case.

It is well known that geometric phases arise only in
time-dependent quantum systems. In the present problem, the
transitions between helicity states on the LRH interfaces, which
is a time-dependent process, results in the geometric phase of
photons. This may be viewed from two aspects: (i) it is apparently
seen in Eq. (\ref{eq7}) that if the frequency parameter $\omega$
in the Hamiltonian (\ref{eq1}) vanishes, then $\dot{\phi}=0$ and
the geometric phase (\ref{eq15}) is therefore vanishing; (ii) it
follows from (\ref{eq2}) that the frequency coefficient $\omega
\left( t\right)$ of Hamiltonian (\ref{eq1}) is

\begin{equation}
\omega \left( t\right) =\zeta \frac{\rm d}{{\rm d}t}p\left(
t\right) =\frac{2c}{nb}\zeta\sum_{k=1}^{\infty }\left[ 1-\left(
-\right) ^{k}\right] \cos \left( \frac{k\pi c}{nb}t\right).
\eqnum{18} \label{eq18}
\end{equation}
Since ${\left| \cos \left( \frac{k\pi c}{nb}t\right)\right|\leq
1}$, the frequency coefficient, the transition rates between
helicity states and the consequent time-dependent phase ( $\varphi
_{\sigma }^{\left( {\rm g}\right) }\left( t\right)+\varphi
_{\sigma }^{\left( {\rm d}\right) }\left( t\right)$ ) greatly
decrease correspondingly as the periodical optical path $nb$
greatly increases. Thus we can conclude that the interaction of
light fields with media on the LRH interfaces gives rise to this
topological quantum phase.

In addition to obtaining the expression (\ref{eq15}) for geometric
phase, we obtain the wavefunction (\ref{eq16}) of photons in the
LRH- optical fiber by solving the time-dependent Schr\"{o}dinger
equation (\ref{eq3}) based on the Lewis-Riesenfeld invariant
theory\cite{Lewis} and the invariant-related unitary
transformation formulation\cite{Gao1,Gao2}. We believe that this
would enable us to consider the propagation of light fields inside
the optical fiber in more detail.

Since we have constructed an effective Hamiltonian (\ref{eq1}) to
describe the time evolution of helicity states of photons, it
should be noted that the method presented here is a
phenomenological description of propagation of lightwave in the
LRH-periodical fiber. This phenomenological description is based
on the assumption that the direction of wave vector ${\bf k}$
becomes opposite nearly instantaneously on the LRH interfaces.
This assumption holds if the periodical length $b$ is much larger
than the wavelength of lightwave in the fiber.
\\ \\
To close this Letter, we conclude with some remarks on the
potential significance of the subject in this Letter:

(i) The obtained geometric phase itself is physically interesting.
Moreover, it is necessary to consider this geometric phase in
discussing the anomalous refraction and wave propagation in
left-handed media ( on the LRH interfaces ).

(ii) Helicity inversion of photons, which is in exact analogy with
the transition operation from $0$ to $1$ in digital circuit, can
be caused due to the electromagnetic interactions on LRH
interfaces, and the time evolution of helicity states is governed
by (\ref{eq16}). We think these processes may possess some
potential applications in information technology and therefore
deserve further investigation.

(iii) It is of physical interest to consider the quantum effects
such as propagation of photons field, polarization of photon
states ( time evolution of photon wavefunction ) and spontaneous
emission decay rate of atoms\cite{Klimov} in left-handed media. In
this Letter, an illustrative example of quantum effects of photons
field resulting from helicity inversion caused by the LRH
interfaces is presented. We hope the present consideration in this
paper would open up new opportunities for investigating more
quantum mechanical properties, phenomena and effects in
left-handed media.
\\ \\
\textbf{Acknowledgements}  I thank X. C. Gao and S. L. He for
useful discussions. This project is supported by the National
Natural Science Foundation of China under the project No.
$90101024$.

\end{document}